 \documentclass[pra,preprint,aps]{revtex4}

\usepackage{amsthm}
\usepackage{amstext}
\usepackage{amssymb}
\newtheorem{thm}{Theorem}
\newtheorem*{thm*}{Theorem}

\begin{document}

\newcommand{\be}{\begin{equation}}
\newcommand{\ee}{\end{equation}}
\newcommand{\ben}{\begin{eqnarray}}
\newcommand{\een}{\end{eqnarray}}
\newcommand{\pp}{\prime}
\newcommand{\nn}{\nonumber}
\newcommand{\dpar}{\partial}

\title{Entropic Upper Bound on Gravitational
Binding Energy}

\author{C. Vignat$^{1}$, A. Plastino$^{2}$ and A.R. Plastino$^{2,3}$}
\address{$^{1}$L.S.S., Supelec, Universit� d'Orsay,
Paris, France, EU  \\
 $^{2}$National University La Plata, CREG-UNLP-CONICET,
 Casilla de Correos 727, 1900 La Plata, Argentina\\
 $^{3}$Instituto Carlos I de F\'{\i}sica Te\'orica
 y Computacional and Departamento de F\'{\i}sica
 At\'omica, Molecular y Nuclear,
 Universidad de Granada, Granada, Spain, EU}

\begin{abstract}

 We prove that the gravitational binding energy $\Omega$
 of a self gravitating system described by a mass density
 distribution $\rho({\bf x})$ admits an upper bound
 $B[\rho(\bf x)]$ given by a simple function of an appropriate,
 non-additive Tsallis' power-law entropic functional
 $S_q$ evaluated on the density $\rho$.
 The density distributions that saturate the entropic
 bound have the form of isotropic $q$-Gaussian distributions.
 These maximizer distributions correspond to the Plummer
 density profile, well known in astrophysics. A heuristic
 scaling argument is advanced suggesting that the entropic
 bound $B[\rho(\bf x)]$ is unique, in the sense that
 it is unlikely that exhaustive entropic upper bounds not
 based on the alluded $S_q$ entropic measure exit. The present
 findings provide a new link between the physics of self
 gravitating systems, on the one hand, and the statistical
 formalism associated with non-additive, power-law entropic
 measures, on the other hand.

\end{abstract}
\maketitle

\section{Introduction}

  The application of non-additive, power law entropic
  functionals (and the concomitant $q$-maxent distributions)
  to the study of self-gravitating systems has attracted
  considerable interest in recent years
  \cite{KKK10,D07,D04,SA04,TS02,TS04a,TS03b,LSS02,SL01,CS05,Y99}. In particular, the
  above mentioned entropic measures and the associated
  $q$-maxent distributions were found to be relevant for the
  study of astrophysical self-gravitating systems such as galaxies and galaxy
  clusters \cite{SF10,SW08,CFSBSR09,HM06,KLK06,L05,MNS04} and of
  other astrophysical problems
  \cite{SCNM05,CNSM09,CSNM08,L02,MSZV10,TPP04},
  such as the statistical characterization of
  the distribution of stellar rotational velocities
  in star clusters, where $q$-maxent distributions were successfully
  applied by Carvalho, do Nascimento, Silva and De Medeiros
  \cite{SCNM05,CNSM09,CSNM08}.

  Compelling evidence for a connection between the physics
  of self-gravitating systems, on the one hand, and the
  statistical formalism associated with Tsallis non-additive
  power law entropy $S_q$ \cite{T88,GT04,T09}, on the other one,
  was  provided by
  the long term, $N$-body numerical simulations performed by
  Taruya and Sakagami \cite{TS03b}. These authors studied
  the behaviour of a stellar system confined within an adiabatic
  wall that evolves from initial conditions not described by
  the Boltzmann-Gibbs distribution and
  found that, before entering the gravothermally unstable
  regime, the evolving self-gravitating system can be described
  remarkably well by a sequence of polytropic distributions
  with a time dependent polytropic index. These polytropic
  distributions have precisely the form of Tsallis'
  $q$-maxent distributions maximizing an $S_q$ entropy
  under the constraints imposed by the total mass and total
  energy of the self-gravitating system, as was first
  pointed out in  \cite{PP93}.
  Taruya and Sakagami also reported numerical evidence
  indicating that the system exhibits the same kind of
  behavior even if the simulations are performed
  without an outer boundary.

  Exact results concerning the equation of state
  within Tsallis thermostatistical framework of
  a two-dimensional gas of gravitationally interacting
  particles were obtained by Sau Fa and Lenzi  \cite{SL01}.
  The Jean's instability of a ``gravitational plasma"
  endowed with a $q$-gaussian velocity distribution
  was analyzed by Lima, Silva, and Santos \cite{LSS02}.
  The gravothermal catastrophe of self-gravitating systems
  characterized by polytropic (that is, $q$-maxent)
  distributions was investigated in detail by Taruya and
  Sakagami (see \cite{TS02,TS04a} and references therein).
  Chavanis and Sire \cite{CS05} showed that the criterion for
  nonlinear dynamical stability in the case of spherical stelar systems
  governed by the Vlasov-Poisson equations exhibits some
  resemblances with a criterion of thermodynamical stability.
  On such basis, Tsallis' functional $S_q$ can be
  regarded as a useful $H$-function that allows for establishing
  a continuous connection between stellar polytropes, on the
  one hand, and isothermal stellar systems on the other one.
  An increasing body of evidence has been reported in recent years indicating
  that the polytropic, $q$-maxent distributions are useful to
  describe dark mater halos of galaxies and galaxy clusters
  \cite{SF10,SW08,CFSBSR09,HM06,KLK06,L05,MNS04,CNSM09,CSNM08}.

  Given a mass density distribution $\rho({\bf x})$ it
  is intuitively clear that its gravitationally binding energy
  $\Omega$ tends to decrease as the system becomes less
  localized. Entropic functionals provide natural quantitative
  measures of the degree of ``delocalization" of the density
  $\rho$. Consequently, one may expect that there is some
  relation between the binding energy $\Omega$ and some
  appropriate entropic functional evaluated on the density
  $\rho$. The aim of the present work is to prove that such a
  relation does indeed exist.

\section{Tsallis Statistics and Gravitation}

 Some of the most important links between the physics
 of self-gravitating systems and the statistical formalism based
 upon Tsallis' entropy come from the study of self-gravitating
 systems described by the Vlasov-Poisson equations \cite{PP93}.
 In order to analyze some important features of galaxies
 and other self-gravitating systems
 it is useful to regard them as non-collisional systems
 constituted by $N$ particles (the stars) interacting
 gravitationally \cite{BT87}. Within this approach, each star moves
 under the effect of the average gravitational potential
 $\Phi({\bf x})$ due to the remaining stars in the galaxy.
 The state of the system is then given by a distribution
 function $F({\bf x},{\bf v},t)$ in position-velocity
 space, where  ${\bf x}$ denotes a star's position vector,
 ${\bf v}$ stands for its velocity vector and $t$ is the
 time. The quantity $F({\bf x},{\bf v},t)d{\bf x} d{\bf v} $
 gives the number of stars having position and velocity
 vectors respectively within the volume elements
 $d{\bf x}$ and $d{\bf v}$ in position and velocity
 spaces.

 The dynamical equation governing the evolution of
 collisionless system is the Vlasov equation

\be \label{vlastelar}
\frac{\partial F}{\partial t}+{\bf v}\cdot\nabla_x F-
\nabla_x \Phi\cdot \nabla_v F=0,
\ee

\noindent coupled to the Poisson equation

\be \label{poisson} \nabla^2 \Phi \, = \, 4 \pi G \rho, \ee

\noindent with $\rho({\bf x},t) = m\, \int F({\bf x},{\bf v},t)
d{\bf v}$, or equivalently to,

\be \label{potvlas} \Phi({\bf x},t) \,  = \, - \, G m \int
\frac{F({\bf x}^{\prime},{\bf v}^{\prime},t)} {|{\bf x} -{\bf
x}^{\prime}|} \, d{\bf x}^{\prime} \, d{\bf v}^{\prime}, \ee

\noindent
which relates the distribution function $F({\bf
x},{\bf v},t)$  with the Newtonian gravitational
potential $\Phi$. Here $G$ denotes Newton's
gravitational constant and $m$ is the mass of
each star (we assume that all stars have the
same mass).

The Vlasov equation can also be cast under the guise $(dF/dt)_{\rm
orb.}=0$. This means that the distribution function $F$ adopts a
constant value along the orbit of any star moving in the
gravitational potential $\Phi$. It is important to stress that the
non-collisional treatment based upon the Vlasov-Poisson system
yields only an approximate description of the dynamics of the
system.  This approximation is, however, a very useful one for
various stellar systems \cite{BT87}. A complete description at the
Newtonian level of the evolution of a self-gravitating
$N$-particle system is given by the complete set of $N$ coupled
Newtonian differential equations of motion corresponding to the
$N$ particles. This approach to the dynamics of self-gravitating
systems is usually investigated by recourse to numerical $N$-body
simulations (see, for instance, \cite{TS03b}).

The total mass,

\be \label{mvlas} M \, = \, m \int \, F({\bf x},{\bf v}) d{\bf
x}\, d{\bf v} \ee

the total energy,

\be \label{evlas} E \, = \, \frac{m}{2} \int  {\bf v}^2 F({\bf
x},{\bf v}) d{\bf x}\, d{\bf v} \, - \, \frac{G m^2}{2} \int
\frac{F({\bf x},{\bf v})F({\bf x}^{\prime},{\bf v}^{\prime})}
{|{\bf x} -{\bf x}^{\prime}|} d{\bf x} \, d{\bf v} \, d{\bf
x}^{\prime} \, d{\bf v}^{\prime},\ee

\noindent
and the total angular momentum,

\be \label{lvlas} {\bf L} \, = \, m \int \, {\bf x} \! \times \!
{\bf v} \, F({\bf x},{\bf v}) \, d{\bf x}\, d{\bf v} \ee

\noindent are preserved under the time evolution determined by the
coupled Vlasov (\ref{vlastelar}) and Poisson (\ref{poisson})
equations.
 The Vlasov-Poisson dynamical system
 (\ref{vlastelar}-\ref{potvlas}) also
 admits a family of conserved quantities
 given by functionals of the form

\be \label{casimir} K[F] \, = \, \int g(F) d{\bf x}d{\bf v}. \ee

\noindent In other words, time dependent solutions $F({\bf x},{\bf
v},t)$ of the Vlasov equation (\ref{vlastelar}) satisfy

\be \label{constcas} \frac{dK}{dt}\, =\, 0. \ee

When one considers different stationary solutions of the
Vlaosv-Poisson system as possible candidates for describing
galaxies or other astrophysical $N$-body systems, it is natural to
ask which is the distribution function $F({\bf x},{\bf v})$ that
maximizes Boltzmann entropy under the constraints given by the
total mass (\ref{mvlas}) and total energy (\ref{evlas}) of the
system. Unfortunately, the formal solution of this variational
problem is given by a distribution function (the isothermal
sphere) exhibiting the physically unacceptable features of having
infinite mass and infinite energy \cite{BT87}. On the other hand,
the optimization of the non additive Tsallis entropic measure
$S_q$ under the same constraints yields the stellar polytropic
sphere distributions which, for a certain range of values of the
entropic parameter $q$, are characterized by a {\it finite} mass
and a finite total energy \cite{PP93}.

The stellar polytropic sphere distributions in position-velocity
space are given by

 \begin{eqnarray} \label{polytropic}
 F({\bf x},{\bf v})=
 F({\tilde \epsilon})=& A (\Phi_0 \, - \,
 {\tilde \epsilon})^{n-3/2}\;\;\;\;&{{\tilde \epsilon}}\leq \Phi_0\\
=&0&{{\tilde \epsilon}}>\Phi_0,\nonumber \end{eqnarray}

\noindent where

\begin{equation} {\tilde \epsilon}=\frac12 {\bf v}^2+\Phi({\bf x}), \label{epsilon}
\end{equation}

 \noindent
denotes the total energy (per unit mass) of a single, individual
star, and $A$, $\Phi_0$, and $n$ (usually referred to as the {\it
polytropic index}) are appropriate constants. After making the
identification of parameters determined by

\begin{equation} \label{quyene}
 \frac{1}{1-q} = n -\frac{1}{2}.
 \end{equation}

\noindent it can by immediately seen that the stellar polytropic
sphere distributions have the $q$-MaxEnt form. That is, they arise
from the maximization of the Tsallis' functional

\be \label{tent} S_q[F] \,=\, \frac{1}{q-1}\, \left(1 \, - \, \int
F({\bf x},{\bf v})^q \, d{\bf x} d{\bf v} \right), \ee

\noindent under the constraints given by the total mass
(\ref{mvlas}) and total energy (\ref{evlas}). Furthermore, the
velocity distribution corresponding to a given space location is a
$q$-Gaussian. The isothermal sphere distribution is recovered in
the limit $n \to\infty$ (corresponding to $q \to 1$). Another
important special case is obtained for $n=5$, which yields the
Schuster sphere distribution \cite{BT87}.

For $n<5$ (and $q<7/9$) the polytropic spheres are systems of
finite total mass and energy. The cut-off exhibited by the
polytropic distribution (\ref{polytropic}) corresponds to what is
known in the field of non extensive thermostatistics  as ``Tsallis
cut-off prescription". The physical origin of this feature of the
stellar polytropic distributions is given by the gravitational
escape velocity corresponding to each value of the radial
coordinate $r$ \cite{BT87}.


\section{An entropic upper-bound on gravitational binding energy}

 The gravitational binding energy $\Omega$ of a mass distribution
 characterized by the density $\rho({\bf x})$ is given by

\be \label{gravine} \Omega \, = \, G \, \int  \, \frac{\rho({\bf
x}) \rho({\bf x}^{\prime})}{|{\bf x} -{\bf x}^{\prime}|} d{\bf x}
\, \, d{\bf x}^{\prime}. \ee

\noindent The system's total gravitational potential energy is
equal to $-\Omega$. In the case of a self-gravitating system
constituted by $N$ particles of mass $m$ and described by a
space-velocity distribution $F({\bf x},{\bf v})$ we have
$\rho({\bf x}) = \int F({\bf x},{\bf v}) d{\bf v}$ and

\be \label{gravinedis} \Omega \, = \, \frac{G m^2}{2} \int
\frac{F({\bf x},{\bf v})F({\bf x}^{\prime},{\bf v}^{\prime})}
{|{\bf x} -{\bf x}^{\prime}|} d{\bf x} \, d{\bf v} \, d{\bf
x}^{\prime} \, d{\bf v}^{\prime}.\ee

It is intuitively clear that the gravitational binding energy
decreases as the system's mass distribution $\rho({\bf x})$
becomes more delocalized. Quantitative measures for this degree of
delocalization are provided by entropic measures evaluated on a
probability density defined as

\be \label{eferho} f({\bf x}) \, = \, \frac{\rho({\bf x})}{M} ,
\ee

\noindent where $M = \int \rho({\bf x}) d{\bf x}$ is the total
mass of the system. Note that $f({\bf x})$ is normalized to $1$.
These considerations suggest that there may be a connection
between the gravitational binding energy $\Omega$, on the one
hand, and an appropriate entropic measure evaluated on the
probability density $p({\bf x})$. Here we are going to prove that,
indeed, such a connection exists and can be established as an
entropic upper bound on the binding energy $\Omega$. We are going
to obtain an upper bound to $\Omega$ given as a monotonic
decreasing function of an appropriate $S_q$ entropy evaluated on
$f({\bf x})$.

We derive the above mentioned entropic upper bound on $\Omega$
from a special case of the Hardy-Littlewood-Sobolev inequality
\cite{Lieb} which reads as follows

\begin{thm}If $q_1=\frac{2D}{2D-\lambda}$ with $0<\lambda<D$ and
$f\in L^{q_1}\left(\mathbb{R}^{D}\right)$ then

\begin{equation}
\label{gravitas}
\Bigl\vert\int_{\mathbb{R}^{D}}\int_{\mathbb{R}^{D}}f \left({\bf
x}\right)f\left({\bf y}\right) \vert {\bf x}- {\bf
y}\vert^{-\lambda}d{\bf x}d{\bf y} \Bigr\vert\le
C\left(D,\lambda\right)\Vert
f\Vert_{q_1}^{2}\label{eq:HLS}\end{equation}

\noindent with the constant

\be C\left(D,\lambda\right)=\pi^{\frac{p}{2}}
\frac{\Gamma\left(\frac{D-\lambda}{2}\right)}{\Gamma
\left(D-\frac{\lambda}{2}\right)}\left(\frac{\Gamma
\left(\frac{D}{2}\right)}{\Gamma
\left(D\right)}\right)^{-1+\frac{\lambda}{D}}.\ee

\noindent The case of equality is reached if and only if

\be \label{maxiq} f\left({\bf x}\right)=A\left(\gamma^{2}+\vert
{\bf x}-{\bf a}\vert^{2}\right)^{-\frac{2D-\lambda}{2}}\ee

\noindent for some $A\in\mathbb{C},$
$\gamma\in\mathbb{R}\backslash\left\{ 0\right\} $ and
$a\in\mathbb{R}^{D}.$\end{thm}

We first remark that this inequality is invariant if ${\bf x}$ and
${\bf y}$ are shifted by ${\bf a}$ so that ${\bf a}$ can be
assumed equal to $0$ without loss of generality. Function
$f\left({\bf x}\right)$ can then be written as

\be \label{efemax} f\left({\bf
x}\right)=A\gamma^{2}\left(1+\frac{\vert {\bf
x}\vert^{2}}{\gamma^{2}}\right)^{-\frac{2D-\lambda}{2}}.\ee

\noindent
 Since moreover inequality (\ref{eq:HLS})
 is homogeneous in $f$, we may
also assume that

\be \label{efenor} \int_{\mathbb{R}^{D}}f\left({\bf x}\right)d{\bf
x}=1.\ee

\noindent The probability density (\ref{maxiq}) is clearly a
$q-$Gaussian distribution. We remark moreover that since
$0<\lambda<D,$

\be \frac{D}{2}<\frac{2D-\lambda}{2}=\frac{1}{1-q_2}<D\ee

\noindent so that

\be
1-\frac{2}{D}<q_2=\frac{2D-\lambda-2}{2D-\lambda}<1-\frac{1}{D}\ee

\noindent The term $\Vert f\Vert_{q_1}^{2}$ in the upper-bound of
(\ref{eq:HLS}) can be recast as,

\be \label{entropicbound} C \Vert f\Vert_{q_1}^{2} = C \left(
\int_{\mathbb{R}^{D}}f^{q_1}\left({\bf x}\right)d{\bf x}
\right)^{\frac{2}{q_1}} = C \left(1+(1-q_1)S_{q_1}[f]
\right)^{\frac{2}{q_1}}, \ee

\noindent which is a monotonically decreasing function of the
entropic functional $S_{q_1}$ evaluated on the density $f$.

The integral appearing in the left hand side of equation
(\ref{gravitas}) can be interpreted as proportional to the binding
energy $\Omega$ of a mass density distribution $\rho({\bf x}) = M
f({\bf x})$ (with total mass $M$) where the interaction of mass
elements $\rho({\bf x_1}) d{\bf x_1}$ and $\rho({\bf x_2}) d{\bf
x_2}$ separated by a distance $r=|{\bf x_2}-{\bf x_1}|$ is
described by a potential function behaving as

\be -\frac{\alpha}{r^{\lambda}} \rho({\bf x_1})\rho({\bf
x_2})d{\bf x_1}d{\bf x_2}, \ee

\noindent where $\alpha$ is a constant characterizing the
``strength" of the above mentioned interaction. The alluded
binding energy is then

\be \label{binding} \Omega \, = \, \alpha M^2
\int_{\mathbb{R}^{D}}\int_{\mathbb{R}^{D}}f \left({\bf
x}\right)f\left({\bf y}\right)\vert {\bf x}- {\bf
y}\vert^{-\lambda}d{\bf x}d{\bf y}. \ee

\noindent Now, combining equations (\ref{gravitas}),
(\ref{entropicbound}) and (\ref{binding}), we get the desired
bound on the binding energy,

\be \label{qbound} \Omega \, \le \, C \alpha M^2  \left( 1 \, + \,
(1-q_1) \, S_{q_1} \!\left[\frac{\rho({\bf x})}{M} \right]
\right)^{\frac{2}{q_1}}. \ee

In the case of the gravitational interaction in $D$-dimensional
space we have $\lambda = D-2$. Consequently, the entropic
parameter $q_1$ corresponding to the entropic bound (\ref{qbound})
is

\be q_1 \, = \, \frac{2D}{D+2}. \ee

In order to get some further physical insight on the particular
structure of the entropic bound $B$, it is instructive to consider
it from the point of view of its scaling behaviour. It is clear that
the binding energy $\Omega$ obeys some simple scaling laws. It seems
reasonable that a physically meaningful entropic upper bound for $\Omega$
should satisfy the same scaling laws as $\Omega$. This requirement is
particularly sensible if the bound is exhaustive, in the sense that
there are density distributions that do achieve the bound. Now, a
natural candidate for an entropy-related functional of $\rho({\bf x})$ satisfying
simple scaling laws is provided by a power-law functional proportional to

\be \label{funqti}
I \, = \, \left[ \int \rho({\bf x})^q \, d{\bf x}  \right]^{\kappa}.
\ee

\noindent
The question is now to find the appropriate
values of the exponents $q$ and $\kappa$.
Let's consider first the scaling transformation
$\rho({\bf x}) \rightarrow  \tilde \rho({\bf x}) = l \rho({\bf x})$,
with $l>0$ a scaling parameter (note that the mass density
$\rho({\bf x})$ is not in general normalized to $1$).
The quantities $\Omega $ and $I$ transform respectively
as $\Omega \rightarrow l^2 \Omega$ and
$I \rightarrow l^{\kappa q} I$, which suggests

\be \label{escala1}
\kappa q =2.
\ee

\noindent
On the other hand, consider the transformation
$\rho({\bf x}) \rightarrow  \tilde \rho({\bf x}) = \rho(l {\bf x})$.
In this case $\Omega$ and $I$ behave as $\Omega \rightarrow l^{-2-D} \Omega$
and $I \rightarrow l^{-\kappa D} I$, suggesting,

\be \label{escala2}
D+2 \, =  \, \kappa D.
\ee

\noindent
Combining (\ref{escala1}) and  (\ref{escala2}) we finally
obtain

\be \label{expo}
q \, = \, \frac{2D}{D+2}, \,\,\,\, {\rm and}   \,\,\,\, \kappa \, = \, \frac{2}{q},
\ee

\noindent
which are precisely the values of the exponents appearing in the bound
(\ref{qbound}). This argument, of course, does not constitute a formal
proof of the bound (the proof, as we already explained, is based on
the Hardy-Littlewood-Sobolev inequality). However, the above scaling
argument suggests that it is unlikely that an entropic bound
for the gravitational binding energy can be found which is not
based on the functional (\ref{funqti}) with the exponents (\ref{expo}).
In other word, the entropic bound (\ref{qbound}) seems to be unique.

\section{$q$-Gaussian maximizers}

 Let us consider now the particular case
 of highest relevance for the study of self gravitating systems,
 which is the one corresponding to three dimensional space ($D=3$)
 and $\lambda = 1$, in which case the integral appearing in
 equation (\ref{binding})) has the standard form of the
 gravitating binding energy of a mass distribution in three
 dimensional  space. In this case we have $q_1 = 6/5$ and
 the entropic upper bound for the gravitational
 binding energy of a system of total mass $M$
 described by a density distribution $\rho({\bf x})$ is,

 \be \label{graventrobound}
 \Omega \, \le \, B[\rho({\bf x})] \, = \, C(3,1)
 \, G M^2 \left( 1 \, - \, \frac{1}{5}
\, S_{6/5} \!\left[\frac{\rho({\bf x})}{M} \right]
\right)^{\frac{5}{3}}.
 \ee

 Let us now consider the density distributions achieving
 the entropic bound (\ref{graventrobound}). Combining equations
 (\ref{eferho}),(\ref{efemax}), and (\ref{efenor}) we can write
 the maximizer mass distributions as

 \be \label{plummer}
 \rho(r) \, = \, \frac{3 M}{4 \pi \gamma^3} \,
 \left(1 + \frac{r^2}{\gamma^2} \right)^{-\frac{5}{2}},
 \ee

 \noindent
 where $r = |{\bf x}|$. Now, equation (\ref{plummer})
 describes the celebrated Plummer density profile, which
 is well known in astrophysics \cite{BT87,P11,D87,AHW74}.
 The parameter $\gamma$ appearing in (\ref{plummer})
 is the ``Plummer radius". It is a scale parameter determining the
 size of the inner ``core" of the mass distribution.

 Plummer's density profile is of considerable theoretical
 interest. It corresponds to a polytropic sphere distribution
 in position-velocity space with polytropic index $n=5$. This
 distribution has the remarkable property of constituting
 the limit case of a polytropic (and a maximum-$S_q$)
 distribution exhibiting a finite total mass and therefore
 being physically plausible. As already mentioned,
 polytropic spheres with $n \le 5$ have finite total mass, while
 those with $n>5$ have infinite mass. Another notable feature
 of the $n=5$ polytropic sphere is that it is a maximum-$S_q$
 distribution in position-velocity space that leads to a density
 distribution (Plummer's distribution) which is itself  a
 $q$-Gaussian (that is, a maximum-$S_q$ distribution).

\section{Conclusions}

 In the present work we obtained an entropic
 upper bound $B[\rho]$ on the gravitational
 binding energy $\Omega$ associated with a
 density mass distribution $\rho({\bf x})$.
 This upper bound is a simple function of an
 appropriate power-law non-additive Tsallis'
 entropy $S_q$ evaluated on $\rho$. In the
 case of the standard, Newtonain gravitational
 interaction in three dimensional space the
 alluded entropy is characterized by the
 entropic parameter $q=\frac{6}{5}$.

 The density distributions achieving the entropic
 bound (that is, satisfying $\Omega = B$)
 exhibit the form of an isotropic $q$-Gaussian
 distribution with $q=\frac{8}{5}$. These
 maximizer distributions coincide with the
 Plummer density profile, well known in
 astrophysics.

 It is well known that many of the important
facts  concerning gravitation are related to simple scaling laws,
Kepler's celebrated third law of planetary motion being
historically the first example \cite{S92}. Here we have provided a
heuristic scaling argument suggesting that the entropic bound
$B[\rho]$ for the gravitational binding energy is unique, with the
specific value of the Tsallis parameter $q_1$ appearing in the
bound determined by the space dimension $D$.

 The present findings constitute intriguing new
 evidence of the relationship that exists between
 the physics of self gravitating systems, on the
 one hand, and the statistical formalism associated
 with non-additive, power law entropic functionals,
 on the other hand.\\

 \noindent
  {\bf Acknowledgments.} This work was partially supported by
  by the Project FQM-2445 of the Junta de Andaluc\'{\i}a
  and by the Grant FIS2008-2380 of the Ministerio de
  Innovaci\'on y  Ciencia, Spain.

\end{document}